\documentclass[onecolumn,useAMS,usenatbib]{mnras}

\usepackage{color}
\usepackage{amsmath}
\usepackage{graphicx}
\usepackage{amssymb}
\usepackage{psfrag}

\newcommand{\beq}{\begin{equation}}
\newcommand{\eeq}{\end{equation}}
\newcommand{\nn}{\nonumber}
\newcommand{\rmd}{\mathrm{d}}
\newcommand{\rmm}{\mathrm{m}}
\newcommand{\rmx}{\mathrm{x}}
\newcommand{\brac}[1]{\left({#1}\right)}
\newcommand{\pd}[2]{\frac{\partial{#1}}{\partial{#2}}}
\newcommand{\td}[2]{\frac{\rmd{#1}}{\rmd{#2}}}

\renewcommand{\div}{\nabla\cdot}

\renewcommand{\bbeta}{{\boldsymbol{\beta}}}
\newcommand{\Lie}{\mathcal{L}}
\newcommand{\half}{{\textstyle{\frac{1}{2}}}}
\newcommand{\sfM}{\mathsf{M}}
\newcommand{\sfL}{\mathsf{L}}
\newcommand{\sfT}{\mathsf{T}}
\newcommand{\sfTheta}{\mathsf{\Theta}}

\newcommand{\red}[1]{{\color{black}{#1}}}

\title[Heat conduction in rotating relativistic stars]{Heat conduction in rotating relativistic stars}
\author[S. K. Lander and N. Andersson]{S. K. Lander${}^1$\thanks{skl@camk.edu.pl},
         N. Andersson${}^2$\\ \\
         ${}^1$Nicolaus Copernicus Astronomical Centre, Polish Academy of Sciences, Bartycka 18, 00-716 Warsaw, Poland,\\
         ${}^2$Mathematical Sciences and STAG Research Centre, University of Southampton, Southampton, SO17 1BJ, U.K.}

\begin{document}

\pagerange{\pageref{firstpage}--\pageref{lastpage}} \pubyear{0000}
\maketitle

\label{firstpage}

\begin{abstract}
In the standard form of the relativistic heat equation used in astrophysics, information
propagates instantaneously, rather than being
limited by the speed of light as demanded by relativity. We show how
this equation nonetheless follows from a more general, causal theory of heat
propagation in which the entropy plays the role of a fluid. In
deriving this result, however, we see that it is necessary to make
some assumptions which are not universally valid: the dynamical
timescales of the process must be long compared with the explicitly
causal physics of the theory, the heat flow must be sufficiently
steady, and the spacetime static. Generalising the heat equation
(e.g. restoring causality) would thus entail retaining some of the terms
we neglected. As a first extension, we derive the heat equation for
the spacetime associated with a slowly-rotating star or black hole,
\red{showing that it only differs from the static result by an
  additional advection term due to the rotation,} and as a
consequence demonstrate that a hotspot on a neutron star will be seen
to be modulated at the rotation frequency by a distant observer.
\end{abstract}



\section{Introduction}

Treating heat propagation in general relativity (GR) is a necessary
ingredient for the quantitative modelling of compact objects. It
becomes important during the collapse of a massive star into either a
neutron star or black hole \citep{misner_sharp64,govender,woosley,sekiguchi}. 
It is needed to understand neutron-star cooling from birth, through the rapid
proto-neutron star phase \citep{burrows_latt} to the slower, secular
evolution of mature neutron stars \citep{vanriper,aguilera}. It is
also important in modelling short-timescale cooling following
outbursts from accreting neutron stars \citep{cumming17}, for
accretion physics around black holes \citep{yuan_narayan,ressler}, and
in the very late stages of a binary neutron-star inspiral \citep{shibata}.

The standard relativistic heat equation is acausal, predicting that
information propagates instantaneously, and thus violating a basic
tenet of relativity that nothing can travel faster than light. It is
now well established that causality can be restored in a natural way, by treating
entropy as a fluid whose dynamics couple to those of the medium
(e.g. the various fluid species of a neutron star) \citep{lopez_and}. Relativistic thermal
dynamics is naturally expressed in this multifluid framework, and can be applied in problems on short
timescales where the finite-speed propagation of entropy is important, or in situations where the
spacetime is highly dynamic -- like the merger of compact objects. In
its full form it is, however, likely to be too computationally complex for many
practical purposes.

Here we show how the usual form of the relativistic heat equation, for
a spherical spacetime, may be recovered from the multifluid framework
in a natural way. Along the way we show that it
is necessary to make a few assumptions -- probably safe ones for
secular processes in a neutron-star or 
black-hole environment, but not necessarily in every astrophysical
situation. This therefore gives a diagnostic of when the familiar relativistic heat equation is not
applicable: if any of these terms is \emph{not} negligible.
Having established the non-rotating result, we generalise our
approach to find the heat equation governing a
slowly-rotating star, keeping the terms of linear order in the
rotation rate (which include frame-dragging). We drop second-order rotational terms (which cause the
spacetime to deviate from sphericity), but note that these have been
explored in the context of neutron-star cooling by \citet{miralles93}
and \citet{negreiros}.

This paper is aimed at an astrophysics audience, and so is intended to
assume no specialist knowledge of the reader. For this reason, we begin with a
brief review of the theory of relativistic thermal dynamics and the
foliation of spacetime for numerical simulations, in order to make our discussion self-contained. We
then derive, in turn, Fourier's law and the heat equation; in each
case beginning with the non-rotating result and then exploring the
effect of slow rotation. We conclude with a \red{discussion of the}
rotational modulation of a neutron-star hotspot seen by a distant observer.


\section{Relativistic thermal dynamics}

\subsection{The problem of causality in heat conduction}
\label{heat_history}

The heat equation is so familiar that it is easy to forget one
conceptually unsatisfactory feature of it. Although this paper
is concerned with heat conduction in general relativity, let us begin in Newtonian
gravity. The heat equation states that a temperature distribution $T$
evolves as:
\beq\label{newt_heat}
\pd{T}{t}=-\frac{1}{C_V}\div{\boldsymbol Q},
\eeq
where $C_V$ is the volumetric heat capacity. The heat flux
${\boldsymbol Q}$ is, in turn, given by Fourier's law:
\beq\label{newt_fourier}
{\boldsymbol Q}=-\kappa\nabla T,
\eeq
where $\kappa$ is the thermal conductivity. Substituting
\eqref{newt_fourier} into \eqref{newt_heat}, however, we see that the
resulting heat equation is parabolic -- and so the characteristic
propagation speed is infinite. Even in Newtonian physics, it is
unreasonable to expect some physical quantity to be transmitted
instantaneously, and efforts to rectify this deficiency in the heat
equation go back decades. A natural reference point is the Cattaneo
equation, in which causality is introduced -- albeit in a
phenomenological manner -- through a time-dependent term in Fourier's
law:
\beq\label{cattaneo}
\mathfrak{t}\pd{{\boldsymbol Q}}{t}+{\boldsymbol Q}=-\kappa\nabla T,
\eeq
where $\mathfrak{t}$ is some small positive number, which can be thought of as a
relaxation timescale for the medium \citep{herrera}. Now plugging this back into equation \eqref{newt_heat}, we have a
hyperbolic equation, known as the telegrapher's equation; the problem
of instantaneous propagation has been removed, but at the expense of
introducing a term not clearly linked to any underlying microphysics.

The relativistic heat equation used for neutron stars and black-hole
spacetimes emerges from the same derivation as in Newtonian gravity,
but with all quantities redshifted, as they are seen by a distant
observer \citep{misner_sharp,thorne67,vanriper}. The causality problem is therefore still present, but is now
even more serious; it is fundamentally unacceptable in GR
for any information to propagate beyond the speed of light,
let alone at infinite speed. The first successful (and not ad-hoc) resolution to the
problem was the Israel-Stewart approach \citep{israel_stew}, which
posits an expansion of the entropy flux through a set of terms which
encode deviations from thermal equilibrium. The various independent
coefficients of this expansion need, however, to be fixed with
additional constraints (either theoretical or experimental). The theory is
thus rather complex, but nonetheless pragmatically
motivated: it allows one to recover a causal heat equation, a
relativistic analogue of the Cattaneo equation.

An alternative starting point is the multifluid (variational) approach
of Carter -- see e.g. \citet{carter89}, or \citet{lopez_thesis} for a fuller account of the
problem and relevant references. In
this, entropy is regarded mathematically as a massless\footnote{more precisely, a
  fluid composed of particles with no \emph{rest} mass} fluid, which
satisfies continuity and Euler equations like any other
fluid. Furthermore, the theory naturally allows for three kinds of
interaction between different fluid species. One can have chemical
reactions describing the creation or destruction of particles --
although, in the case of entropy, the second law of thermodynamics
dictates that the reaction rate must be non-negative. Secondly,
dissipation is introduced through a series 
of scattering terms, which include scattering of the usual fluid particles with
the entropy `particles' (e.g. phonons). The third, and least
familiar, interaction is \emph{entrainment}. This is a non-dissipative
coupling between two fluid species, which depends on the relative
velocity between the two.  In normal fluid dynamics, the momentum and
velocity of a fluid are parallel; in the presence of entrainment the
two can be misaligned, a phenomenon sometimes known as the Andreev-Bashkin
effect \citep{andbash,alpar84} in the context of mixed Fermi liquids.

Within the multifluid framework, it is certainly \emph{permitted} to allow
for entrainment between the entropy fluid and other species, but
the physical reason for doing so is not initially obvious. Indeed, in the
original incarnation of his multifluid model, Carter dropped these
terms for simplicity. It was quickly seen, however, that the resulting theory suffered
from serious instabilities \citep{olson_his}, casting doubt on the usefulness of the
framework until \citet{priou} showed that the theory was
indeed stable once the entrainment terms were restored. In fact, the
resulting model can be shown to be equivalent to the Israel-Stewart theory up
to first order in deviations from thermal equilibrium \citep{lopez_and}.

Despite its superficially obscure nature, entrainment between the
entropy fluid and the matter fluid(s) in a system is also the key ingredient
which allows one to derive a Cattaneo-type form of Fourier's law, and thus restore causality to heat conduction within the
multifluid approach -- both in Newtonian gravity and general
relativity \citep{lopez_thesis}. Because it acts over timescales much shorter than those of many
relativistic astrophysics problems, it might appear to be irrelevant
for these. Let us sound two notes of
caution, though. Firstly, the standard heat equation will not be adequate for
modelling every hot relativistic system -- especially not those
involving dynamic spacetimes or processes with short characteristic
timescales. Secondly, even when it should be valid, there is
still a risk that the neglect of entrainment could lead to
instabilities, as for Carter's original approach. For these reasons,
whilst we will indeed drop entropy entrainment during the course of
deriving the heat equation, we will keep it for the first steps, to
show where the explicitly causal terms lie.

\subsection{The equations of entropy dynamics}

Our starting point will be \citet{and17}, hereafter AHDC, who
use a general-relativistic multifluid formalism in which
entropy becomes, mathematically, just another fluid. Thermal dynamics
are then described by one scalar entropy equation, and one vector
entropy-momentum equation. We begin, as AHDC
do, in geometrised units where $G=c=1$ (so that all quantities have
dimensions with powers of length alone), then restore these factors
afterwards. AHDC utilise the standard `3+1 split', in which
4-dimensional spacetime is foliated into a nested set
of 3-dimensional spacelike hypersurfaces threaded by a set of timelike
worldlines which cut through each hypersurface perpendicularly; see e.g. \citet{TM82}.
In what follows, we will denote spacetime quantities using indices
$a,b$ (taking values from $0$ to $3$); and will use the indices $i,j,k$
(taking values from $1$ to $3$) to denote quantities restricted to the
hypersurfaces. We will give only a minimal description of the 3+1 split,
and refer the reader to the notes of \citet{gour07} for a detailed, pedagogical description.

An observer travelling along a timelike worldline (often called an
\emph{Eulerian observer}) experiences
proper time $\tau$ and has a 4-velocity ${\boldsymbol N}$ (in index notation, $N^a$) given by $\rmd/\rmd\tau$. The
relationship between an observer's proper time, and the `global time' $t$
measured by an observer at infinity, is encoded in the \emph{lapse}
$\alpha$, defined as
\beq
\alpha\equiv\td{\tau}{t}
\eeq
along a worldline.

The notion of time variation may be expressed in terms of the 3+1
split, as the Lie derivative along the 4-vector
\beq
{\boldsymbol t}=\alpha{\boldsymbol N}+{\boldsymbol{\beta}}
\ \ \ \textrm{or}\ \ \ t^a=\alpha N^a+\beta^a,
\eeq
using a coordinate basis for the second expression.
Here $\boldsymbol{\beta}$ is the `shift vector': a 3-vector living in a spacelike
hypersurface, so that
\beq
\beta^a N_a=0.
\eeq
We may therefore denote the shift vector $\beta^i$ rather than
$\beta^a$, when it does not result in mismatched indices within the
same expression. The shift vector may be arbitrarily specified, aside
from the restriction of being spatial.
After this general summary section, we
will specify to stationary and axisymmetric spacetimes,
i.e. spacetimes with timelike and azimuthal Killing vectors. The
arbitrariness of the shift vector discussed above is then very useful, because we may
define it so that $\alpha{\boldsymbol N}+{\boldsymbol{\beta}}$ is equal to the timelike
Killing vector (the natural notion of time in this context).

Now, from the lapse and shift we can split the spacetime, writing $x^a=(t,x^i)$, so that the
line element reads
\beq\label{line_elem}
dl^2 = g_{ab}dx^a dx^b= -(\alpha^2-\beta_i\beta^i)dt^2 + 2\beta_i dx^i dt + \gamma_{ij}dx^i dx^j,
\eeq
where $g_{ab}$ is the spacetime metric and $\gamma_{ij}$ the 3-metric of the spatial hypersurfaces.

Returning to derivatives, we may use the linearity of the Lie
derivative to show that
\beq
\partial_t=\Lie_{\boldsymbol t}=\Lie_{\alpha{\boldsymbol N}+{\boldsymbol{\beta}}}
         =\Lie_{\alpha{\boldsymbol N}}+\Lie_{\boldsymbol{\beta}}.
\eeq
Since the Lie derivative of a covariant 4-vector $v_a$ along some other
  4-vector $u^a$ is given by
\beq \label{Lie_4vec}
\Lie_{\boldsymbol u} v_a=u^b v_{a,b}+v_b u^b_{,a},
\eeq
we see that for a scalar $h$:
\beq\label{Lie_scalar}
\Lie_{\boldsymbol N}h=\frac{1}{\alpha}(\partial_t-\Lie_\bbeta) h.
\eeq
The same is also true for a covariant 3-vector $p_i$. To see this,
first note that since $N_a$ is timelike, $N_a q^a=0$ for a 3-vector
$q^i$, and therefore $0=N_a q^a=g_{ab} N^b g^{ac} q_c=N^b q_b$. Using
this result and equation \eqref{Lie_4vec}, we then find that:
\beq
\Lie_{\alpha{\boldsymbol N}}p_i=\alpha N^b p_{i,b}+p_j(\alpha N^j)_{,i}
 =\alpha N^b p_{i,b}+\alpha p_j N^j_{,i}=\alpha\Lie_{\boldsymbol N}p_i,
\eeq
and so
\beq \label{Lie_3vec}
\Lie_{\boldsymbol N}p_i=\frac{1}{\alpha}(\partial_t-\Lie_\bbeta) p_i.
\eeq
In addition to the Lie derivative, we will briefly use the covariant
derivative for spacetime quantities, denoting it with a subscript
semicolon; and will extensively use the covariant derivative projected
into the hypersurfaces, $D_i$ (see AHDC for more details). We recall its form, in components, when
acting on co- and contra-variant 3-vectors, $p_i$ and $q^i$ respectively:
\begin{align}
\label{D_cov}
D_j p_i &= p_{i,j}-\Gamma^k_{\ ji}p_k ,\\
D_j q^i &= q^i_{,j}+q^k\Gamma^i_{\ jk} ,
\label{D_con}
\end{align}
where the Christoffel symbols $\Gamma^i_{\ jk}$ (associated with the
projected covariant derivatives) encode the difference
between covariant and partial derivatives of a 3-vector which arises from the
curvature of spacetime. In our context we only have derivatives within
a hypersurface, for which these symbols involve derivatives of the
3-metric $\gamma_{ij}$ only:
\beq
\Gamma^i_{\ jk}
 = \half\gamma^{il}(\gamma_{jl,k}+\gamma_{lk,j}-\gamma_{jk,l}).
\eeq
Note that the 3-metric is diagonal in the cases we consider,
i.e. $\gamma^{il}=0$ for $i\neq l$.

\red{Within the 3+1 framework, the 4-velocity $u_\rmx^a$ of any fluid
species $\rmx$ is split as follows:
\beq\label{vel_split}
u_\rmx^a=W_\rmx(N^a+\hat{v}_\rmx^a),
\eeq
where $\hat{v}_\rmx^a$ is the spacelike 3-velocity of the fluid within a
hypersurface, as seen by an Eulerian observer, and
\beq
W_\rmx=(1-\hat{v}_\rmx^2)^{-1/2}
\eeq
is the Lorentz factor associated with the $\rmx$-fluid.
As mentioned earlier, thermal dynamics in the multifluid formalism is described quite
naturally by treating entropy as a massless `fluid'
\citep{prix04}. Then, the entropy flux $n^a_s$ under the 3+1 split is given by
\beq
n^a_s=n_s u_s^a=W_s n_s(N^a+\hat{v}_s^a)=\hat{n}_s(N^a+\hat{v}_s^a)=\hat{s}(N^a+\hat{v}_s^a),
\eeq
where a hat over a symbol (e.g. $\hat{n}_s$) refers to a quantity
measured by an Eulerian observer; unaccented quantities in this paper are
either universal or referred to a fluid frame. Here we have recognised that the number density $\hat{n}_s$ of the entropy
fluid -- measured by an Eulerian observer -- is the entropy
\emph{density} $\hat{s}$}. Like all fluids, the entropy has a continuity
equation associated with it:
\beq \label{ent_cont}
(n^a_s)_{;a}=\Gamma_s\geq 0,
\eeq
where $\Gamma_s$ is the entropy production rate. The only thing which
distinguishes this continuity equation from the usual form is
that the production rate must be non-negative in this case, by the second
law of thermodynamics. We shall call equation
\eqref{ent_cont} the entropy equation.

The other fundamental quantity is the entropy 4-momentum:
\beq
\mu_a^s=\hat{T} N_a+S^s_a,
\eeq
where we have identified the time component of the 4-momentum as the
temperature $\hat{T}$, and denoted the entropy 3-momentum by $S_a^s$ (which
is spacelike; it satisfies $N^a S_a^s=0$).
Again, in analogy with ordinary fluids, the non-conservation of the 
entropy 4-momentum is due to chemical reactions and frictional
processes \red{(most naturally measured by an observer in the fluid frame
with some 4-velocity $u_a$)}, as expressed in the momentum equation:
\beq \label{entmom_general}
(N^a+v_s^a)(\mu^s_{a;b}-\mu^s_{b;a})+\Gamma_s\mu_a^s
 = \Gamma_s T(N_a+v_a^s)
      +\sum\limits_{{\textrm{m}}\neq s}\mathcal{R}^{s\textrm{m}}(\delta_a^b+v_s^b u_a)(v_b^{\textrm{m}}-v_b^s),
\eeq
which may be obtained by combining equations (64), (65) and (72) from
AHDC with equation \eqref{vel_split} of this paper. Note that here,
and later, we will generally refer to all fluid species other than
the entropy as matter fluids (assumed to move together), with index $\mathrm{m}$. Equation
\eqref{entmom_general}, being an intermediate algebraic step, is not
the usual guise in which the momentum equation appears. Nonetheless,
we can identify the familiar features by thinking of the spacetime
coordinate with index $0$ as time and coordinates $1,2,3$ as
spatial. Then, we see that equation \eqref{entmom_general} features a time-derivative
of the momentum as well as its divergence -- representing the
notion from Newton's second law that the imbalance of forces on a
fluid source its net acceleration -- together with various terms which
describe the transfer of momentum into or out of the entropy-fluid component
by dissipation or reactions.

As discussed in section \ref{heat_history}, entropy entrainment is
crucial for constructing a causal theory of thermal dynamics. Although
we will not retain this effect later on, it is
instructive to see where it features in these fundamental equations. In principle the entrainment term
could then be propagated through the rest of our derivation of Fourier's
law to yield an explicitly causal final heat equation. It appears in
the definition of the entropy 3-momentum (see AHDC equation (96)):
\beq\label{StoT}
S_s^i=\hat{T}\hat{v}_s^i+\sum\limits_{\rmm\neq s}\mathcal{A}^{s\rmm}\hat{n}_\rmm(\hat{v}_\rmm^i-\hat{v}_s^i),
\eeq
from which it is easy to see that entrainment allows for the entropy momentum
and velocity to be misaligned. The timescale on which a
coefficient $\mathcal{A}^{s\rmm}$ couples the matter and entropy
dynamics should, therefore, be some analogue of the phenomenological
relaxation timescale $\mathfrak{t}$ from the Cattaneo equation
\eqref{cattaneo} \citep{lopez_thesis}.

We have introduced the entropy \eqref{ent_cont} and entropy momentum
\eqref{entmom_general} equations as they emerge from the multifluid
framework. A major part of the work of AHDC was to rewrite and
simplify these to 3+1 forms closer to those used in numerical
relativity. We refer the reader to their paper for derivations, and
here simply present the required results. Firstly, the entropy
equation may be rewritten as [AHDC equation (136)]:\red{
\beq \label{orig_ent}
\partial_t(\gamma^{1/2}\hat{s})
+D_i\Bigg\{\alpha\gamma^{1/2}\left[\frac{Q^i}{\hat{T}}+\brac{\hat{v}^i-\frac{\beta^i}{\alpha}}\hat{s} \right]\Bigg\}
=\alpha\gamma^{1/2}\Gamma_s,
\eeq
where $\gamma=g^{ab}\gamma_{ab}$ is the determinant of the spatial
3-metric, $\hat{v}^i$ the 3-velocity of a fluid-frame observer, and
where we have defined a more physically-motivated quantity -- the heat
flux -- as
\beq \label{Q_def}
Q^i\equiv \hat{s}\hat{T} (\hat{v}_s^i-\hat{v}^i)
  = \hat{s}S^i_s -\hat{s}\hat{T}\hat{v}^i- \hat{s}\hat{T}\sum\limits_{\rmm\neq s}\mathcal{A}^{s\rmm}\hat{n}_\rmm(v_\rmm^i-v_s^i).
\eeq
This quantity -- analogous to the electric current -- depends on the relative
flow of two fluids with respect to one another. In the case of the
heat flux the two fluid species are matter and entropy. Note that the
vanishing of $Q^i$ does not imply the vanishing of $S_s^i$, nor
vice-versa: the heat flux vanishes when a fluid observer measures
the entropy velocity as zero, whereas the entropy momentum vanishes when an
\emph{Eulerian} observer sees zero entropy velocity. AHDC make use of
a linear-drift approximation -- that the drift of the entropy fluid
with respect to the matter fluid should be slow -- which implies that
$Q^2\sim(\hat{v}_s-\hat{v})^2\ll Q$. This means that in expressions featuring
different powers of $Q$, we only need to retain those terms of the
lowest power.
We now turn to the expression for the entropy momentum equation given
in\footnote{\red{The above equation corrects an error from AHDC. Specifically, between their equations (78) and
  (80) the extrinsic curvature tensor $K_{ij}$ erroneously appears in
  place of the correct term, involving the trace $K$ of this
  tensor. The incorrect term is then propagated through their later
  derivations, including into their form of the entropy momentum
  equation (139).} } AHDC equation (139), which features both $Q^i$ and $S_s^i$
explicitly. Using equation \eqref{Q_def} then allows us to eliminate the entropy 3-momentum
in favour of the heat flux and entropy entrainment terms.  
If we now \emph{neglect} entrainment, the result is:
\begin{align}
\partial_t[\gamma^{1/2}(Q_i+\hat{s}\hat{T}\hat{v}_i)] 
 &+D_j\Big[\alpha\gamma^{1/2}\frac{Q^j}{\hat{s}\hat{T}}(Q_i+\hat{s}\hat{T}\hat{v}_i) \Big]
   +D_j\Big[\gamma^{1/2}\brac{\alpha\hat{v}^j-\beta^j}(Q_i+\hat{s}\hat{T}\hat{v}_i) \Big]\nn\\
   +& \hat{s}D_i(\alpha\gamma^{1/2}\hat{T})
  - \brac{\frac{Q^j}{\hat{T}}+\hat{s}\hat{v}^j}D_i\left[\alpha\gamma^{1/2}\brac{\frac{Q_j}{\hat{s}}+\hat{T}\hat{v}_j}\right]
 = \gamma^{1/2}\left[\alpha\mathcal{F}^s_i-\alpha K \hat{s} S^s_i+(Q_j+\hat{s}\hat{T}\hat{v}_j)D_i\beta^j\right],
\label{orig_entmom}
\end{align}
}where the entropy entrainment terms -- were they to be restored -- would feature together with every
instance of the heat flux, except for the two contravariant $Q^j$
terms on the left-hand side. In equation \eqref{orig_entmom} we have
introduced two new quantities. The first is the trace of the extrinsic curvature
\beq\label{extrin}
K=\gamma^{ij} K_{ij} = \frac{1}{2\alpha}\gamma^{ij}\brac{ -\gamma_{ij,t} + \beta^k\gamma_{ij,k}
                                                    + \gamma_{kj}\beta^k_{,i} + \gamma_{ik}\beta^k_{,j} },
\eeq
which is related to how a hypersurface curves within the spacetime in
which it is embedded; and the second is a resistive term $\mathcal{F}^s_i$
which encodes the transfer of momentum away from the
entropy component due to collisions with other particle species. For clarity, let us
consider only entropy-matter particle scattering, and
neglect any additional resistive mechanisms (e.g. Joule heating from
magnetic-field decay). Under these assumptions, the collision term
given by equation (142) of AHDC reduces to:
\beq \label{scattering}
\mathcal{F}^s_i = -\frac{\mathcal{R}}{WsT}[Q_i+W^2\hat{v}_i \hat{v}^j Q_j]
\eeq
and the creation rate term given by AHDC equation (143) becomes:
\beq \label{creation}
\Gamma_s=\frac{\mathcal{R}}{W^2 s^2 T^3}[Q^2+W^2(\hat{v}_j Q^j)^2],
\eeq
where we have defined two quantities: the Lorentz factor
\beq
W\equiv (1-\hat{v}^2)^{-1/2},
\eeq
and the sum of the different entropy-matter
scattering coefficients
\beq
\mathcal{R}\equiv \sum\limits_{\rmm\neq s}\mathcal{R}^{\rmm s}.
\eeq
Note that the quantities in equations \eqref{scattering} and
\eqref{creation} are related to the microphysics of the system,
naturally performed in the fluid frame, and so do not have hats.
The system of equations for relativistic thermal dynamics summarised
above, from AHDC, is appropriate for nonlinear evolutions in
dynamic spacetimes using a 3+1 foliation of spacetime (even though, by
equation \eqref{orig_entmom}, the system has already ceased to be
valid for studying processes on timescales short compared with the
medium's thermal relaxation). However, there are many problems for which this
is unduly general. In what follows, we will explore how these
equations simplify in one typical astrophysical setting.

\subsection{Rotating systems and observers}
\label{rot_observer}

Let us consider heat conduction in the spacetime associated with a
central object -- either a star or a black hole -- rotating uniformly at rate
$\Omega$ \red{with respect to a distant observer. Its 4-velocity is therefore
\beq
u^a=\frac{1}{\alpha}(\delta^a_t+\Omega\delta_\varphi^a).
\eeq
}  We will assume that the object is isolated, so that the
spacetime remains stationary, and that the rotation is sufficiently slow that
we may drop the second-order rotational terms which lead to deviations
from spherical symmetry (e.g. the oblateness induced by rotation, in
Newtonian or relativistic stars).

From here onwards, we will begin to restore the
suppressed $G$ and $c$ factors (i.e. the dimensions) to the expressions
from AHDC. By doing so, we will
immediately be able to identify combinations of unfamiliar quantities
from the multifluid framework with familiar transport
properties, using dimensional analysis. Converting from geometrised to
physical units is not especially simple, and so we establish the
requisite conversions systematically, by beginning with the most
fundamental quantities. To start with, it is natural to identify
the spatial coordinates for our system with globally-defined
spherical polar coordinates, $(ct,r,\theta,\varphi)$. The line
element, with geometrising factors unsuppressed, is then
\beq \label{line_elem2}
dl^2 = -e^{2\Phi/c^2}c^2 dt^2-2\frac{\omega}{c} r^2\sin^2\theta d\varphi c dt
             + e^{2\Lambda/c^2} dr^2 + r^2 d\theta^2 + r^2\sin^2\theta d\varphi^2,
\eeq
where $\Phi=\Phi(r)$ and $\Lambda=\Lambda(r)$ are the two metric
functions, and $\omega$ is the angular velocity at which frames
near the central object are dragged with respect to an observer at
infinity.

It is very useful to generalise the Newtonian notion of a corotating
observer to the relativistic one of a zero-angular-momentum observer
(ZAMO) \citep{bardeen70,bardeen72}; these are the Eulerian observers
for rotating systems. A ZAMO has a local
rotational velocity of zero, and the mathematical description of
physical processes is at its least complex with respect to such an
observer. More specifically, one has a set of equations governing the
physics of the system (e.g. a neutron star), in terms of a
globally-defined system of coordinates. The local coordinate system of
a ZAMO is encoded in a tetrad of orthonormal basis vectors; projecting
the globally-defined equations onto a ZAMO's orthonormal tetrad
results in a greatly simplified description of the physics.
From this perspective we may regard $\omega$ as the variation, with respect to global time $t$,
of the ZAMO's azimuthal coordinate:
\beq
\omega = \td{x^{\hat{3}}}{x^{0}}=\td{\hat{\varphi}}{t}.
\eeq
This quantity can be shown to be a function of the radial coordinate
alone \citep{hartle67}. It is given by the equation
\beq \label{hartle}
\frac{1}{r^3}\td{ }{r}\brac{r^4 e^{-(\Phi+\Lambda)/c^2}\td{\bar{\omega}}{r}}
 + 4\td{ }{r}\brac{e^{-(\Phi+\Lambda)/c^2}}\bar{\omega}=0,
\eeq
where
\beq\label{baromega}
\bar{\omega}=\Omega-\omega
\eeq
is the angular velocity of a fluid element as seen by a ZAMO.
Now, comparing the two line elements \eqref{line_elem} and \eqref{line_elem2} -- which means identifying the ZAMO 4-velocity
$u^a$ with the normal $N^a$ -- we see that the lapse and shift are
given by:
\beq
\alpha= e^{\Phi/c^2}\ ,
\ \beta^i =-\frac{\omega}{c}\delta^i_{\varphi}.
\eeq
It is now clear that the $\beta_i\beta^i=\beta^2$ term in the 3+1 line
element, equation \eqref{line_elem}, corresponds to second-order
rotational corrections in our problem, and therefore may be neglected.\red{
We now also know that the fluid 3-velocity, as seen by a ZAMO, is
\beq
\hat{v}^i=e^{-\Phi/c^2}\bar\omega\delta_\varphi^i
\eeq
and so all Lorentz factors reduce to unity, since
\beq
\brac{1-\frac{\hat{v}^2}{c^2}}^{-1/2}=\brac{1-e^{-2\Phi/c^2}\frac{\bar\omega^2}{c^2}}^{-1/2}
 = 1+\frac{1}{2!}e^{-2\Phi/c^2}\frac{\bar\omega^2}{c^2}+\dots\approx 1,
\eeq
using our assumption of slow rotation.}

A brief calculation using equation \eqref{extrin} shows that the
extrinsic curvature $K_{ij}$ vanishes for a non-rotating star,
but in a rotating system has one independent non-zero component,
which in physical units is (using primes to denote derivatives with
respect to $r$):
\beq \label{ext_curv}
K_{r\varphi}=-\half e^{-\Phi/c^2}r^2\sin^2\theta\omega'(r) =K_{\varphi r}
\eeq
since $K_{ij}$ is symmetric \citep{gour07}. \red{To get the trace
$K$ of this quantity, however, we must contract it with the 3-metric
$\gamma^{ij}$, which is diagonal -- and therefore $K=0$.}

\subsection{Physical quantities and their dimensions}
\label{dimensions}

Let us pause to discuss the quantities involved in the equations
for thermal dynamics and also their dimensions, which will help in the
interpretation of physical quantities later on in our derivations. We denote by
$\sfM,\sfL,\sfT$ and $\sfTheta$ the dimensions of mass, length, time and
temperature. The physical dimensions of the two basic thermal quantities, $T$ and
$s$, are
\beq
[T_\mathrm{phys}]=\sfTheta\ ,\ [s_\mathrm{phys}]=[\mathcal{S}]\sfL^{-3}=\sfM\sfL^2\sfT^{-2}\sfTheta^{-1}\times\sfL^{-3}
=\sfM\sfL^{-1}\sfT^{-2}\sfTheta^{-1},
\eeq
where $\mathcal{S}$ is the true entropy (i.e. not per unit volume).
In standard geometrised units for temperature-independent problems,
one sets $G=c=1$, and all quantities have dimensions which are powers
of $\sfL$. It is possible to extend this to relativistic thermal dynamics by
setting $G=c=k_B=1$, where $k_B$ is the Boltzmann constant;
e.g. one divides entropy by $k_B$, so that its geometrised form is
dimensionless. Since no algebra from AHDC involved factors of $k_B$
anyway though, their equations are the same in either system of geometrised
units. Accordingly, we will proceed with the simpler $G=c=1$ system,
allowing the dimensions of each quantity to contain powers of both $\sfL$ and $\sfTheta$.

In a $G=c=1$ unit system, $T$ remains the same, but we must multiply $s$ by
a prefactor combination of $G$ and $c$ to eliminate $\sfT$ and $\sfM$:
\beq
s_\mathrm{geom}=\frac{G}{c^4}s_\mathrm{phys}
\implies [s_\mathrm{geom}]=\sfL^{-2}\sfTheta^{-1}.
\eeq
Next, all velocities in geometrised units are of the form
\beq
v^i_\mathrm{geom}=\frac{v^i_\mathrm{phys}}{c}
\eeq
and so are dimensionless.
The heat flux $Q^i$ in geometrised units is therefore
\beq
Q^i_\mathrm{geom}=s_\mathrm{geom}T(v_s^i-v^i)_\mathrm{geom}
       =\frac{G}{c^5}s_\mathrm{phys}T(v_s^i-v^i)_\mathrm{phys}
\eeq
and has dimensions
\beq
[Q^i_\mathrm{geom}]=\sfL^{-2}.
\eeq
If we divide through by the geometrising prefactor we get
\beq
[Q^i_\mathrm{phys}]=[c]^5 [G]^{-1}[Q^i_\mathrm{geom}]=\sfM\sfT^{-3},
\eeq
which are indeed the expected physical units for heat flux.
Next, because our physical coordinates are $(ct,r,\theta,\varphi)$, time
derivatives must contain a $1/c$ factor, which is suppressed in
geometrised units:
\beq
(\partial_t)^\mathrm{geom}=\frac{1}{c}(\partial_t)^\mathrm{phys}.
\eeq
This means a time derivative in geometrised units has dimensions $\sfL^{-1}$.
From \eqref{orig_ent} we then see that
\beq
[\Gamma_s^\mathrm{geom}]=[(\partial_t s)^\mathrm{geom}]
\eeq
and so
\beq
\Gamma_s^\mathrm{geom}=\frac{G}{c^5}\Gamma_s^\mathrm{phys}.
\eeq
Using equation \eqref{orig_entmom} we can determine that in
geometrised units
\beq
[\mathcal{F}_i^s]=[s D_i \alpha T]=\sfL^{-2}\sfTheta^{-1}\times\sfL^{-1}\times \sfTheta
 = \sfL^{-3},
\eeq
and so
\beq
(\mathcal{F}_i^s)^\mathrm{geom}=\frac{G}{c^4}(\mathcal{F}_i^s)^\mathrm{phys}.
\eeq
Finally, by equation \eqref{scattering} we have
\beq
[\mathcal{R}]=[\mathcal{F}_i^s][s][T][Q_i]^{-1}
 = \sfL^{-3}\sfL^{-2}\sfTheta^{-1}\sfTheta(\sfL^{-2})^{-1}=\sfL^{-3}.
\eeq
Again restoring the suppressed geometric prefactors, we have
\beq
\mathcal{R}_\mathrm{geom}=\frac{G}{c^3}\mathcal{R}_\mathrm{phys}
\eeq
i.e.
\beq
[\mathcal{R}_{phys}]=\sfM\sfL^{-3}\sfT^{-1}.
\eeq

\section{Fourier's law}

The standard form of Fourier's law, equation \eqref{newt_fourier}, relates the heat flux to the
temperature gradient, with the thermal conductivity as the constant of
proportionality. If heat
conduction is different in different directions -- e.g. due to the
effect of a magnetic field -- one needs to replace the scalar conductivity with a
tensorial one ${\boldsymbol\kappa}$, so that ${\boldsymbol Q}=-{\boldsymbol\kappa}\cdot\nabla T$
\citep{UY80}. As discussed in section \ref{heat_history},
relativistic forms of Fourier's law and the heat equation are by no
means novel (e.g. \citet{vanriper,aguilera}), and date back to at least the 1960s
\citep{misner_sharp,thorne67}. However, these have been generalised from their
Newtonian counterparts in a simple way, essentially by replacing
flat-space quantities in the derivations of these equations by redshifted ones (i.e. the locally-measured
value of some quantity in a spherical spacetime needs to be multiplied by a factor of
$e^{\Phi/c^2}$ to yield the value seen by a distant observer). This
makes it difficult to see how causality could be restored to the heat
equation, or what new terms would appear in a more complex
relativistic system.  Here, by contrast, we aim to derive relativistic
forms of Fourier's law and the heat equation from the manifestly
causal multifluid formalism, in which the route to generalising the
model is also clear. We will first recover the expected equation for
Fourier's law in a non-rotating star, then will consider the problem
with terms at first order in the rotation.

Firstly, let us recall from AHDC the results:
\beq
[(-g)^{1/2}]_{;a}=(\alpha\gamma^{1/2})_{;a}=0\ \ \ ,\ \ \ 
 D_i(\alpha\gamma^{1/2})=\partial_i(\alpha\gamma^{1/2})-\Gamma^j_{ji}\gamma^{1/2}=0.
\eeq
In addition
\beq
(\gamma^{1/2})_{,t}=(\alpha\gamma^{1/2})_{,t}=0
\eeq
by the stationarity of the spacetime we consider; recall that we are
not allowing for the spacetime itself to evolve here.
Given these, we may pull these quantities out of the covariant and time
derivatives, and cancel them. The entropy momentum equation
\eqref{orig_entmom} in geometrised form then reduces to:\red{
\beq
\frac{1}{\alpha}\partial_t(Q_i+\hat{s}\hat{T}\hat{v}_i) 
  +\brac{\hat{v}^j-\frac{\beta^j}{\alpha}} D_j(Q_i+\hat{s}\hat{T}\hat{v}_i)
  +D_j\Big[\frac{Q^j}{\hat{s}\hat{T}}(Q_i+\hat{s}\hat{T}\hat{v}_i) \Big]
   +\hat{s}D_i \hat{T}
  -\brac{\frac{Q^j}{\hat{T}}+\hat{s}\hat{v}^j}D_i\brac{\frac{Q_j}{\hat{s}}+\hat{T}\hat{v}_j}
 = \mathcal{F}^s_i+\frac{1}{\alpha}(Q_j+\hat{s}\hat{T}\hat{v}_j)D_i\beta^j,
\label{entmom2}
\eeq
where we have used the fact that both our shift vector and the
(rigidly-rotating) fluid flow are divergence-free,
$D_j\beta^j=D_j\hat{v}^j=0$, and that $K=0$.}
Comparing equations \eqref{Lie_3vec} and \eqref{D_cov}, we see that
the first two terms of the above equation \eqref{entmom2} may be rewritten as follows:
\beq\label{LieN_rewrite}
\frac{1}{\alpha}(\partial_t-\beta^j D_j)(Q_i+\hat{s}\hat{T}\hat{v}_i)
 +\hat{v}^j D_j (Q_i+\hat{s}\hat{T}\hat{v}_i)
 = \Lie_{\boldsymbol N} (Q_i+\hat{s}\hat{T}\hat{v}_i)
    +\frac{1}{\alpha} (Q_j+\hat{s}\hat{T}\hat{v}_j)\beta^j_{,i}
    +\frac{1}{\alpha}\beta^j\Gamma^k_{\ ji}(Q_k+\hat{s}\hat{T}\hat{v}_k)
    +\hat{v}^j D_j (Q_i+\hat{s}\hat{T}\hat{v}_i).
\eeq
Next, we expand and rearrange the covariant derivative from the
right-hand side of equation \eqref{entmom2}:
\beq\label{covRHS_rewrite}
\frac{1}{\alpha}(Q_j+\hat{s}\hat{T}\hat{v}_j)D_i\beta^j
= \frac{1}{\alpha} (Q_j+\hat{s}\hat{T}\hat{v}_j)\beta^j_{,i}
    +\frac{1}{\alpha}(Q_k+\hat{s}\hat{T}\hat{v}_k)\beta^j\Gamma^k_{\ ji},
\eeq
where we have relabelled the indices and used the fact that the
Christoffel symbols are symmetric in the lower two indices.
Inserting these last two results into the entropy momentum equation (still in
geometrised form), a number of terms cancel and we are left with:
\beq \label{lots_of_vQ}
\frac{1}{\hat{s}}\Lie_{\boldsymbol N}(Q_i+\hat{s}\hat{T}\hat{v}_i)
 +\frac{1}{\hat{s}}\hat{v}^j D_j Q_i + \frac{1}{\hat{s}} D_j(Q^j\hat{v}_i)
 -\frac{Q^j}{\hat{s}\hat{T}}D_i(\hat{v}_j\hat{T})
 +\hat{v}^j D_i\brac{\frac{Q_j}{\hat{s}}}
 +D_i \hat{T} = \frac{1}{\hat{s}}\mathcal{F}^s_i,
\eeq
where we have also made use of the result $Q^2\ll Q$ (which follows
from the linear-drift approximation; see AHDC) in order to drop terms quadratic
in the heat flux. \red{The Lie derivative of $\hat{s}\hat{T}\hat{v}_i$ may
be rewritten using a couple of results derived later in this
paper. Rather than breaking the flow here to discuss these here, we
refer the reader forward to section \ref{heat_section}. Specifically, we use equations
\eqref{ent_eqn} and \eqref{Lie_s_T} together with the assumption of rigid, slow
rotation, to find that
\beq
\Lie_{\boldsymbol N}(\hat{s}\hat{T}\hat{v}_i)
= \brac{1+\frac{\hat{s}}{C_V}}\hat{T}\hat{v}^i
      \left[ \Gamma_s-D_j\brac{\frac{Q^j}{\hat{T}}} \right].
\eeq
We see that this term, like many others in equation
\eqref{lots_of_vQ}, is proportional to $\hat{v}Q$. But since
$\hat{v}\sim\hat{v}_s$ (linear-drift approximation), we know that
\beq
\hat{v}Q\propto \hat{v}(\hat{v}-\hat{v}_s)\propto\hat{v}^2,
\eeq
and so for consistency with our slow-rotation approximation all of
these terms should be neglected.} The entropy momentum equation in
physical units then reduces to
\beq \label{general_entmom}
\frac{1}{c^2 \hat{s}}\Lie_{\boldsymbol N}Q_i+D_i \hat{T}=-\frac{\mathcal{R}}{s^2 T}Q_i,
\eeq
where we have restored the suppressed $G$ and $c$ factors and plugged
in the explicit form of the scattering term \eqref{scattering}.

\subsection{Non-rotating limit}

We can understand equation \eqref{general_entmom} by looking at the
limit in which $\omega\to 0$. This corresponds to no frame dragging,
and therefore no rotation in relativistic gravity. We find that:
\beq \label{fourier_norot0}
\frac{1}{c^2\alpha \hat{s}}\partial_t Q_i+D_i \hat{T} = -\frac{\mathcal{R}}{s^2 T}Q_i.
\eeq
Comparing with the usual form of Fourier's law \eqref{newt_fourier}, let us
identify the following quantity as the heat conductivity $\kappa$:
\beq \label{kappa_0}
\kappa \equiv \frac{s^2 T}{\mathcal{R}}.
\eeq
We can use dimensional analysis as a consistency check of this
definition, using results from section \ref{dimensions}; we find that
$[\kappa]=\sfM\sfL\sfT^{-3}\sfTheta^{-1}$,
which are indeed the expected physical units.
Now, equation \eqref{fourier_norot0} becomes:
\beq \label{fourier_norot1}
\frac{1}{c^2\alpha \hat{s}}\partial_t Q_i+D_i \hat{T} = -\frac{1}{\kappa} Q_i.
\eeq
This is not quite what we want though: the usual form of Fourier's law
has no time dependence, whereas here we find a time derivative of the heat
flux. Equally though, this factor does \emph{not} render equation
\eqref{fourier_norot1} the kind of stable, causal expression
resulting from the inclusion of the entropy entrainment -- despite its
superficial resemblance to equation \eqref{cattaneo} (see, e.g.,
\citet{lopez_thesis}). This is because the
prefactor of $\partial_t Q_i$ in equation \eqref{fourier_norot1} is not
`tuneable' and has no connection with the medium through which heat
propagates. Only in the limit of Newtonian dynamics, where $Q/c\propto
v_s/c\to 0$, does the time-derivative term vanish automatically.

At this point we see that in GR we can recover the standard Fourier's law
only if the heat flow is approximately steady on a thermal
timescale. More precisely, we want the timescale $\tau_Q$ for
variations in the heat flux to satisfy:
\beq
\tau_Q\gg \frac{\kappa}{c^2\alpha \hat{s}},
\eeq
but the $c^2$ factor means that this assumption ought to be quite
safe -- at least for processes on secular timescales. Finally then, we
can reach the desired result:
\beq \label{fourier_norot}
D_i \hat{T} = -\frac{1}{\kappa} Q_i.
\eeq

\subsection{Slow rotation}

Having thus identified the heat conductivity from the Newtonian limit
of the entropy momentum equation, let us return to the general case, 
equation \eqref{general_entmom}. For the same reasons as in the
non-rotating case, we again want to assume the heat flow is steady
over dynamical timescales. The
natural notion of time variation in the foliation framework, however,
is not with respect to global time, $\partial_t$, but rather a local
expression given by the Lie derivative $\Lie_{\boldsymbol N}$ along the normal
to the hypersurfaces:
\beq\label{Lie_N}
\mathcal{L}_{\boldsymbol N}=\frac{1}{\alpha}(\partial_t-\Lie_\bbeta ).
\eeq
We therefore neglect this term from the left-hand
side of equation \eqref{general_entmom}, substituting in equation
\eqref{kappa_0} for $\kappa$, to find that
\beq\label{fourier_slowrot}
D_i \hat{T}=-\frac{1}{\kappa}Q_i,
\eeq
as in the non-rotating case. Thus, Fourier's law will only differ from
its form in a non-rotating star for the case of rapid rotation.

\section{The heat equation}
\label{heat_section}

We used the entropy momentum equation \eqref{orig_entmom} to derive Fourier's law,
above. Now we use the corresponding scalar entropy equation \eqref{orig_ent} to derive
the heat equation. As for the derivation of Fourier's law, we begin by
taking the $\gamma^{1/2}$ and $\alpha\gamma^{1/2}$ factors out of the
covariant and time derivatives and cancelling the former, leaving us with:
\beq \label{ent_eqn1}
\partial_t \hat{s}+\alpha D_i\brac{\frac{Q^i}{\hat{T}}}-(\alpha\hat{v}^i-\beta^i)D_i \hat{s}
  = \alpha\Gamma_s,
\eeq
where we have again used the fact that $D_j\beta^j=D_j\hat{v}^j=0$.
\red{Within the 3+1 approach, it is natural to re-express equation
\eqref{ent_eqn1} in terms of the Lie derivative along the timelike
normal ${\boldsymbol N}$:
\beq \label{ent_eqn}
\Lie_{\boldsymbol N} \hat{s}-\hat{v}^i D_i \hat{s}
 +D_i\brac{\frac{Q^i}{\hat{T}}} = \Gamma_s,
\eeq
where the shift vector is given by the ZAMO rotational frequency, and
this is the form of the entropy equation we will use. Note,
however, that the equation takes an even simpler form when expressed
in terms of a `material derivative'
 $\Lie_{\boldsymbol u}\equiv\alpha^{-1}(\partial_t-\Omega\delta^i_\varphi D_i)$
with respect to a corotating fluid observer (rather than a ZAMO):
\beq
\Lie_{\boldsymbol u} \hat{s} +D_i\brac{\frac{Q^i}{\hat{T}}}  = \Gamma_s.
\eeq
}
Note that these various forms of the entropy equation are all the same in both physical and geometrised
units; in the geometrised case each of the terms has a factor of
$G/c^5$ suppressed, and so these may be cancelled. Now, by virtue of
our slow-rotation approximation (meaning that $\hat{v}^2$ terms are
negligible and $W=1$), the entropy creation rate from equation
\eqref{creation} may be simplified to
\beq
\Gamma_s = \frac{\mathcal{R}}{s^2 T^3} Q^2=\frac{Q^2}{\kappa T^2},
\eeq
which, together with the second law of thermodynamics, implies that
$\mathcal{R}>0$. Although the entropy creation rate is proportional to
$Q^2$ and therefore small compared with $Q$, it cannot be neglected, since it is not manifestly smaller than
other terms in the entropy equation.
The next part of the strategy will be to rewrite terms involving
derivatives of the entropy, using the first law of thermodynamics:
\beq
\rmd \mathcal{U}
 = T\rmd \mathcal{S} - P\rmd V +\sum\limits_{\textrm{x}\neq s}\mu_{\textrm{x}}\rmd \mathcal{N}_{\textrm{x}},
\eeq
where $\mathcal{S}$ and $\mathcal{U}$ are the true entropy and
internal energy (as opposed to the quantities per unit volume we use
elsewhere), $P$ is pressure, $V$ volume and $\mathcal{N}_\rmx$ the number of
$\rmx$-particles. Let us assume the total
number $\mathcal{N}$ of matter particles is conserved, so that
$\rmd\mathcal{N}=0$; then, for a system with a single species of
matter particle, the third term becomes $\mu\rmd\mathcal{N}=0$. In the
case of multiple particle species, we get a similar result if we make
the additional assumption of chemical equilibrium; then
$\rmd\mathcal{N}_\rmx=0$ for each species, and so the first law becomes:
\beq\label{firstlaw}
\rmd \mathcal{U} = T\rmd \mathcal{S} - P\rmd V.
\eeq
From equations \eqref{ent_eqn} and \eqref{firstlaw} we will derive the
heat equation -- first for a non-rotating star, to show that we
recover the expected result, and then for the case of slow rotation.

\subsection{Non-rotating limit}

Taking the time derivative of the first law \eqref{firstlaw} per unit
volume, we find that
\beq \label{dUdt}
\pd{U}{t}\Big|_V = T \pd{s}{t}\Big|_V,
\eeq
where $U$ is internal energy per unit volume, and we recall that $s$
is entropy per unit volume.
Now expand the left-hand side of this expression using the chain rule:
\beq \label{firstlaw_norot}
\pd{U}{t}\Big|_V = \pd{U}{T}\Big|_V \pd{T}{t}\Big|_V \equiv C_V\pd{T}{t}\Big|_V,
\eeq
using the definition of $C_V$. Note that we do not have an
additional $\partial U/\partial s$ term from applying the chain rule
in equation \eqref{firstlaw_norot}, because $s$ and $T$ are
thermodynamic conjugate pairs, and we may regard $U$ as being a
function of either variable. Now comparing the above two equations, we can eliminate $\partial_t \hat{s}$
in favour of  $\partial_t \hat{T}$ in the entropy equation \eqref{ent_eqn1}, which for
constant-volume processes becomes:
\beq
C_V \partial_t \hat{T}+\alpha \hat{T} D_i\brac{\frac{Q^i}{\hat{T}}} = \frac{Q^2}{\kappa T^2}.
\eeq
Use of the product rule on this expression then gives:
\beq \label{ent_dstodt}
C_V \partial_t \hat{T}+\alpha D_i Q^i -\alpha\frac{Q^i}{\hat{T}} D_i \hat{T}= \frac{Q^2}{\kappa T^2}.
\eeq
Rewriting the third term with Fourier's law \eqref{fourier_norot}, we have
\beq
\alpha\frac{Q^i}{\hat{T}} D_i \hat{T}=\alpha\frac{Q^2}{\kappa T},
\eeq
since $\hat{T}=WT\approx T$, which cancels with the right-hand-side term of
\eqref{ent_dstodt}. Then, by additionally using
$D_i(\alpha\gamma^{1/2})=D_i(\gamma^{1/2})=0$ on the resulting
expression, we can manipulate it as follows:
\beq
C_V\alpha\gamma \partial_t \hat{T}=\alpha^2\gamma D_i(\kappa D^i \hat{T})
=D_i[\alpha\gamma^{1/2} \kappa D^i(\alpha\gamma^{1/2} \hat{T})]
=\gamma D_i[\alpha \kappa D^i(\alpha \hat{T})],
\eeq
so giving us
\beq
C_V \partial_t(e^\Phi \hat{T}) =D_i[e^\Phi \kappa D^i(e^\Phi \hat{T})]
\textrm{\ \ \ or\ \ \ }  C_V \partial_t(e^\Phi \hat{T}) =\nabla\cdot[e^\Phi \kappa \nabla(e^\Phi \hat{T})],
\eeq
which is the usual way in which the relativistic heat equation is
presented, in terms of the redshifted temperature $e^\Phi \hat{T}$.

\subsection{Slow rotation}

For rotating stars the logic used in deriving the heat equation is the same, but because the spacetime is
no longer static we again need to generalise the notion of time
derivative to be the Lie derivative along the normal vector (see
equation \eqref{Lie_N}). With this concept, equation \eqref{dUdt} from the static case
generalises to:
\beq
\mathcal{L}_{\boldsymbol N} U\Big|_V = T \mathcal{L}_{\boldsymbol N} s\Big|_V,
\eeq
and the chain rule gives
\beq
\mathcal{L}_{\boldsymbol N} U\Big|_V = \pd{U}{T}\Big|_V \mathcal{L}_{\boldsymbol N} T\Big|_V
 =C_V \mathcal{L}_{\boldsymbol N} T\Big|_V.
\eeq
Combining these last two equations, as for the non-rotating case, gives
\beq\label{Lie_s_T}
\mathcal{L}_{\boldsymbol N} s = \frac{C_V}{T}\mathcal{L}_{\boldsymbol N} T.
\eeq
We now return to the entropy equation \eqref{ent_eqn}, expanding the
covariant derivatives with the product rule to
show that:
\beq \label{preheat}
\Lie_{\boldsymbol N} \hat{s}-\hat{v}^i D_i \hat{s}
 +D_i\brac{\frac{Q^i}{\hat{T}}} = \frac{\mathcal{R}}{s^2 T^2} Q^2=\frac{1}{\kappa T} Q^2.
 \eeq
\red{Now plugging Fourier's law (equation \eqref{fourier_slowrot}) and the definition
of $\Lie_{\boldsymbol N}$ into equation \eqref{preheat}, we find that:
\beq\label{heat_s}
\hat{T}(\Lie_{\boldsymbol N}-\hat{v}^i D_i)\hat{s}+D_i Q^i = \frac{Q^2}{\kappa T}+\frac{1}{\hat{T}}Q^i D_i \hat{T}=0,
\eeq
again exploiting $T\approx\hat{T}$. The second half of the first term above corresponds to the entropy being advected
with the rotating fluid flow, and simplifies to:
\beq
\hat{v}^i D_i \hat{s}=\frac{1}{\alpha}\pd{\hat{s}}{\varphi}=\frac{C_V}{\alpha\hat{T}}\pd{\hat{T}}{\varphi},
\eeq
by the same arguments which gave equations \eqref{dUdt} and \eqref{firstlaw_norot}.
Using this result and equation \eqref{Lie_s_T} to replace $\hat{s}$ with $\hat{T}$
in equation \eqref{heat_s} then yields the heat equation for a
rotating star,
\beq
C_V\brac{\mathcal{L}_{\boldsymbol N}-\frac{\Omega}{\alpha}\pd{ }{\varphi}} \hat{T}=-D_i Q^i,
\eeq
or, by expanding $\mathcal{L}_{\boldsymbol N}$ and using Fourier's law,
\beq  \label{heatrot_general}
C_V\brac{\pd{\hat{T}}{t}-\bar\omega\pd{\hat{T}}{\varphi}} =D_i(\kappa D^i \hat{T})
\textrm{\ \ or\ \ }
C_V\brac{\pd{\hat{T}}{t}-\bar\omega\pd{\hat{T}}{\varphi}} =\div(\kappa\nabla\hat{T}).
\eeq
Note that in the limit of zero diffusion ($\kappa=0$), the equation above
describes a temperature distribution depending on the quantity
$(\varphi+\bar\omega t)$, as expected: the temperature moves forwards with
angular velocity $\bar\omega=\Omega-\omega$ with respect to a ZAMO.}

We have shown that one recovers the expected heat equation if only the
lowest-order rotational corrections are kept. Next, we briefly make
contact with astrophysics, by considering one particular physical situation
with direct relevance for neutron-star observations.

\section{Rotational modulation of a neutron-star hotspot}

There are a number of observations of hotspots on neutron stars which show
modulation in time.  In some cases the frequency at which they are
modulated is the only way to determine their rotation
frequency \citep{stroh97}. Some neutron stars accreting in low-mass X-ray binary
systems produce X-ray bursts due to thermonuclear burning in the
neutron-star ocean, and in some cases the bursts display almost
coherent oscillations -- typically in the range $\sim 300-600$
Hz \citep{watts12}. These
oscillations are believed to be related either to modes of the
neutron-star ocean \citep{heyl04}, or to an essentially confined
hotspot \citep{cumming00} -- though there are challenges with either
model in explaining the small frequency drifts of $\sim 1$ Hz seen in
the burst oscillations. In either case, the characteristic frequency
is believed to be due to rotational modulation, which implicitly
assumes that an observer at infinity really sees a hotspot on the
stellar surface moving at the rotation rate. Let us close by
considering how reasonable this is for rapidly-rotating neutron stars.

We have seen that for slow rotation -- retaining only terms linear in
the rotational frequency -- the heat equation takes the form of equation
\eqref{heatrot_general}. The frame-dragging term on the left-hand side
of this equation tells us that a hotspot fixed on the surface of a
neutron star will be rotationally-modulated for a ZAMO at a frequency
of $\Omega-\omega$. However, the ZAMO itself moves at
$+\omega$ with respect to an observer at infinity -- so, for this
latter observer the hotspot is indeed rotationally modulated at the
expected frequency of $\Omega$. \red{We have not studied more rapidly
rotating stars here, for which terms at order $\Omega^2$ should be
retained, but there is no reason for the modulation of a hotspot to be affected by
the oblate shape induced by order-$\Omega^2$ terms. The other class of
higher-order terms is those proportional to $Q\Omega$, neglected by us just before
equation \eqref{general_entmom}. Since they describe a coupling
of the stellar rotation with the heat flux, it is indeed possible that
these terms would alter a hotspot's motion on a rapidly-rotating star.

In conclusion, within the slow-rotation approximation a hotspot
is indeed seen modulated at exactly the stellar rotation rate. Even
the most rapidly-rotating neutron stars known are still rather `slow', in the
sense that their rotational kinetic energy is small compared with the
gravitational binding energy, so our heat equation is probably
adequate for describing all known neutron stars (with the caveat that
we had to make various physical assumptions to derive it). There
remains the possibility, however, that higher-order couplings will
produce a hotspot rotating at a slightly different rate from the star.}

\section{Summary}

We have shown how the standard form of the relativistic heat equation
follows from a causal theory of heat propagation, in which the entropy
is treated as a fluid. \red{At first order in rotation, the only correction
is the expected one that the temperature is advected with the fluid
flow. We have not studied the second-order problem, but speculate that
it may result in coupling between the heat flow and the rotation.}

One benefit in having performed our detailed derivation 
is that it demonstrates how the heat equation relies
implicitly on various assumptions which will \emph{not} be safe in all
astrophysical situations. The dynamical timescale for the problem at
hand must be long compared to the thermal relaxation time of the
medium, or equivalently the timescale on which entrainment couples the
entropy and matter fluids. Equally, the heat flux must
be approximately steady over short timescales. Even in situations
where these conditions are met, we should be mindful that a thermal
evolution relying on the standard heat equation could still suffer genuine
instabilities connected with the neglect of entropy entrainment.

It is not so surprising that the usual form of the heat equation may
be recovered in the limiting case described above, where thermal
information propagates almost `instantaneously' -- compared with any
fluid dynamics -- and where the spacetime is stationary. For more
dynamical situations, however, the equations presented here are not
applicable -- and a future goal for the numerical simulations of hot
relativistic systems should be to evolve the entropy dynamics directly.

\section*{Acknowledgements}

SKL acknowledges support from the European Union's Horizon 2020
research and innovation programme under the Marie Sk\l{}odowska-Curie
grant agreement No. 665778, via fellowship UMO-2016/21/P/ST9/03689 
of the National Science Centre, Poland.
Both authors acknowledge support from STFC via grant number ST/M000931/1, and from
NewCompStar (a COST-funded Research Networking Programme). We thank
the referee for an insightful report, which helped considerably in improving this paper.

\label{lastpage}

\end{document}